\documentclass[12pt]{article}
\usepackage{epsfig}
\usepackage{graphicx}
\usepackage{color}
\usepackage{amsmath,amsfonts,amsthm, amssymb,latexsym}
\usepackage{subfigure}

\def\beq{\begin{equation}}
\def\eeq#1{\label{#1}\end{equation}}
\def\eeqn{\end{equation}}

\def\beqa{\begin{eqnarray}}
\def\eeqa#1{\label{#1}\end{eqnarray}}
\def\eeqan{\end{eqnarray}}

\let\bar=\overbar

\def\Dslash{\not{\hbox{\kern-4pt $D$}}}
\def\dslash{\not{\hbox{\kern-2pt $\del$}}}

\def\msb{{\bar{\ssstyle M \kern -1pt S}}}

\usepackage{fancyhdr,graphicx}
\def\Title#1{\begin{center} {\Large {\bf #1} } \end{center}}

\begin{document}

\Title{Many-body forces, isospin asymmetry and dense hyperonic matter}


\begin{raggedright}

{R.O. Gomes$^{1,2}$~~V. Dexheimer$^{3}$~~S. Schramm$^{2}$~~and C.A.Z. Vascconcellos$^{1}$\\
\bigskip
$^{1}$ Instituto de F\'isica, Universidade Federal do Rio Grande do Sul, Porto Alegre, RS 91501-970 Brazil
\\
\bigskip
$^{2}$Frankfurt Institute for Advanced Studies, 60438 Frankfurt am Main, Germany
\\
\bigskip
$^{3}$Department of Physics, Kent State University, Kent, OH 44242, USA
}

\end{raggedright}

\section{Introduction}

The equation of state (EoS) of asymmetric nuclear matter at high densities is a key topic for the description of matter inside neutron stars.
The determination of the properties of asymmetric nuclear matter, such as the symmetry energy ($a_{sym}$) and the slope of the symmetry energy ($L_0$)
at saturation density, has been exaustively studied in order to better constrain the nuclear matter EoS.
However, differently from symmetric matter properties that are reasonably constrained, the symmetry energy and its slope still large uncertainties in 
their experimental values.
Regarding this subject, some studies point towards small values of the slope of the symmetry energy 
~\cite{Steiner2012, Lattimer2013, Li2013}, while others suggest rather higher values 
~\cite{Chen2005, Tsang2012, Cozma2013, Wang2014, Sotani2015}. Such a lack of agreement raised a certain debate in the scientific community. 
In this paper, we aim to analyse the role of these properties on the behavior of asymmetric hyperonic matter.
Using the formalism presented in Ref. \cite{ROGomes2014}, which considers many-body forces contributions in the meson-baryon coupling,
we calculate the EoS of asymmetric hyperonic matter (section 2) and apply it to describe hyperonic matter and hyperon stars (section 3).

\section{Formalism}

In this work, we use the formalism developed in Ref. \cite{ROGomes2014} that takes into account many-body contributions to the nuclear force 
via a parameterized derivative coupling for the mesons.
Besides the $\sigma$ and $\omega$ mesons present in RMF models, we also consider the $\varrho$ and $\delta$ mesons, that 
allow a better description of the properties of asymmetric matter, as well as the
strange $\phi$ meson, that is important for the description of hyperon interactions. 
Also, as we are foccusing on stiffer EoS, we neglect the strange $\sigma^*$ meson, since it introduces more attraction to the system.
The Lagrangian density of the model is (see Ref. \cite{ROGomes2014} for more details):

\small
\begin{equation} \label{lagrangiana_zen2}
\begin{split}
\mathcal{L}&= {\sum}_{b}\overline{\psi}_{b}\left[\gamma_{\mu}\left(i\partial^{\mu}-g_{\omega b }\omega^{\mu} -g_{\phi b }\phi^{\mu}
-g_{\varrho b} I_{3b} \varrho_{3}^{\mu} \right)
-\left(m_b - g_{\sigma b}^{*}\sigma -g_{\delta b}^{*}I_{3b}\delta_{3}\right)\right]\psi_{b}
\\& +\frac{1}{2}\left(\partial_{\mu}\sigma\partial^{\mu}\sigma-m_{\sigma}^{2}\sigma^{2}\right)
+\frac{1}{2}\left(\partial_{\mu}\delta_{3}\partial^{\mu}\delta_{3}-m_{\delta}^{2}\delta_{3}^{2}\right)
+\frac{1}{2}\left(-\frac{1}{2}\omega_{\mu\nu}\omega^{\mu\nu}+m_{\omega}^{2}\omega_{\mu}\omega^{\mu}\right)
\\&+\frac{1}{2}\left(-\frac{1}{2}\phi_{\mu\nu}\phi^{\mu\nu}+m_{\phi}^{2}\phi_{\mu}\phi^{\mu}\right)
+\frac{1}{2}\left(-\frac{1}{2}\varrho^{3}_{\mu\nu}\varrho_{3}^{\mu\nu}+m_{\varrho}^{2} \varrho^{3}_{\mu}.\varrho_{3}^{\mu}\right)
+{\sum}_{l}\overline{\psi}_{l}\gamma_{\mu}\left(i\partial^{\mu}-m_{l}\right)\psi_{l}.
\end{split}
\end{equation}
\normalsize

The subscripts $b$ and $l$ represent the baryon octet ($n$, $p$, $\Lambda^0$, $\Sigma^-$, $\Sigma^0$, $\Sigma^+$, $\Xi^-$, $\Xi^0$)  
and lepton ($e^-$, $\mu^-$) degrees of freedom, respectively. 
The Dirac Lagrangian density for baryons and leptons are in the first and last terms, while
the other terms represent the Lagrangian densities of the mesons: 
a Klein-Gordon Lagrangian density for the scalar $\sigma$ and $\delta$ fields and a Proca Lagrangian density for the
vector $\omega$, $\varrho$, and $\phi$ fields. 
The baryon isospin component in the $z-$direction in represented by $I_{3b}$, which correspond to the terms that take the isospin asymmetry into account

As we use the so called scalar version of the model in this work, the many-body forces contributions are introduced only for the 
scalar meson-baryon coupling constants, present in the effective mass term of Equation ~\ref{lagrangiana_zen2}. 
The  definition of the meson-baryon couplings are:
\begin{equation}\label{coupling_zen}
g_{\sigma b}^{*}\equiv \left(1+\frac{g_{\sigma b}\sigma+g_{\delta b} I_{3b}\delta_{3}}{\zeta\,m_{b}}\right)^{-\zeta} g_{\sigma b},
\quad g_{\delta b}^{*}\equiv \left(1+\frac{g_{\sigma b}\sigma+g_{\delta b} I_{3b}\delta_{3}}{\zeta\,m_{b}}\right)^{-\zeta} g_{\delta b}
\end{equation}
where the coupling introduces nonlinear contributions, controlled through the $\zeta$ parameter of the model and 
the effective mass of baryons have the following expression: 
$ m^*_{b \zeta} \equiv \, m_b - g_{\sigma b}^* \sigma - g_{\delta b}^* I_{3b}\delta_{3}$.

Also, as we allow hyperonic degrees of freedom to appear at high densities,
we describe their coupling constants to the $\omega$ and $\phi$ vector mesons using 
SU(6) spin-flavor symmetry ~\cite{Dover1985,Schaffner1993}, as follows:
\begin{equation}
\label{hys}
\frac{1}{3}g_{\omega N}= \frac{1}{2}g_{\omega \Lambda}=\frac{1}{2} g_{\omega \Sigma}=  g_{\omega \Xi}, \qquad
-\frac{2\sqrt{2}}{3} g_{\omega N}= 2 g_{\phi \Lambda}= 2 g_{\phi \Sigma}= g_{\phi \Xi},
\end{equation}
and also assume an isospin scaling for the coupling to the isovector mesons:
\begin{equation}
g_{\varrho N}=\frac{1}{2} g_{\varrho \Sigma}=g_{\varrho \Xi}, \quad g_{\varrho \Lambda} = 0, \qquad
g_{\delta N}=\frac{1}{2} g_{\delta \Sigma}=g_{\delta \Xi}, \quad g_{\delta \Lambda} = 0.
\end{equation}
Finaly, the coupling to the sigma meson is obtained by fitting the potential depths of 
the hyperons in nuclear matter ~\cite{Glendenning1991,Schaffner1992}
for the values $U_{\Lambda}^N =-28\, \mathrm{MeV}$, $U_{\Sigma}^N=+ 30\, \mathrm{MeV}$, and $U_{\Xi}^N= -18\, \mathrm{MeV}$ ~\cite{SchaffnerBielich2000}.
In what follows, we vary the values of the symmetry energy and its slope and analyze their effects on the properties of hyperonic matter
and hyperon stars. 

\section{Results and Discussion}

Using the mean field approximation, we calculate the EoS of the model from the components of the stress-energy tensor,
imposing a saturation density of $\rho_0= 0.15 \,\mathrm{fm}^{-3}$, and a binding energy per baryon of $B/A = -15.75\,\mathrm{MeV}$.
Also, by the fitting of the mesons' coupling constants, we apply the EoS to describe the following nuclear matter properties at saturation: 
effective mass of the nucleon $m_N^*$, compressibility modulus $K_0$, symmetry energy $a^0_{sym}$, and symmetry energy slope $L_0$.
These quantities and the respective values of the coupling constants are displayed in Table~\ref{table:Table_coupling}, for different parametrizations.

\begin{table}
\centering{}%
\scriptsize
\begin{tabular}{ccccccccc}
\hline 
\tabularnewline
$\zeta$ & $m^*_n/m_n$ &  $K_0$ & $a_{sym}^0$ &  $L_0$ & $(g_{\sigma N}/m_{\sigma})^2$ & $(g_{\omega N}/m_{\omega})^2$& $(g_{\varrho N}/m_{\varrho})^2$ & $(g_{\delta N}/m_{\delta})^2$  \\
   &  &  (MeV) & (MeV) &  (MeV) & ($\mathrm{fm}^{2}$) & ($\mathrm{fm}^{2}$)& ($\mathrm{fm}^{2}$)& ($\mathrm{fm}^{2}$) \\
   \hline 
   \tabularnewline

  0.040 & 0.66 & 297 & 25 & 76  & 14.51  & 8.74 & 2.56  & 0.38 \tabularnewline
  0.040 & 0.66 & 297 & 25 & 90  & 14.51  & 8.74 & 6.53  & 6.50 \tabularnewline
  0.040 & 0.66 & 297 & 33 & 100 & 14.51  & 8.74 & 4.74  & 0.38 \tabularnewline
  0.040 & 0.66 & 297 & 33 & 115 & 14.51  & 8.74 & 6.93  & 0.66 \tabularnewline
\\
  0.049 & 0.68 & 272 & 25 & 74  & 13.99  & 8.14&  2.53  & 0.16 \tabularnewline
  0.049 & 0.68 & 272 & 25 & 90  & 13.99  & 8.14 & 7.20  & 7.56 \tabularnewline
  0.049 & 0.68 & 272 & 33 & 100 & 13.99  & 8.14 & 5.31  & 1.11 \tabularnewline
  0.049 & 0.68 & 272 & 33 & 115 & 13.99  & 8.14 & 9.65  & 8.02 \tabularnewline
  \\
  0.059 & 0.70 & 253 & 25 & 73  & 13.44  & 7.55 & 2.75  & 0.34 \tabularnewline
  0.059 & 0.70 & 253 & 25 & 90  & 13.44  & 7.55  & 7.78  & 8.54 \tabularnewline
  0.059 & 0.70 & 253 & 33 & 100 & 13.44  & 7.55  & 5.84  & 1.82 \tabularnewline
  0.059 & 0.70 & 253 & 33 & 115 & 13.44  & 7.55  & 10.24  & 9.00 \tabularnewline

  \hline 
\end{tabular}\medskip{}
\caption{\label{table:Table_coupling} Nuclear matter properties for different parametrizations of the model (different $\zeta$'s): 
effective mass of the nucleon, compressibility modulus, symmetry energy, symmetry  energy slope and ($\sigma,\,\omega,\,\varrho,\,\delta$) coupling constants.}
\end{table}           
     
\normalsize

In addition, there is still debate in the literature regarding the values of the asymmetric matter properties at saturation ~\cite{Lattimer:2012}.
For this reason, we vary their values in ranges of $a^0_{sym}$ between $25 - 33\,\mathrm{MeV}$ ~\cite{Tsang2012} and 
$L_0$ between $73-100 \,\mathrm{MeV}$ ~\cite{Chen2005,Lopes:2014wda,Bizarro2015,Oertel2015}, 
in order to verify their effects on the behavior of hyperonic matter at high densities. 

As discussed in the previous section, the effective mass of baryons depends on the $\zeta$-parameter which, consequently, is also
affected by the nonlinear contributions that come from the many-body forces coupling. 
Hence, each parametrization has different particle populations, as one can see in Ref. \cite{ROGomes2014}.
Furthermore, as the isospin asymmetric properties ($a_{sym}^0$ and $L_0$) have direct impact on the coupling constants of the isovector mesons ($\varrho$, $\delta$),
they also have effect on the particles population.

It is important to emphasize that, for a fixed value of the symmetry energy, 
the lowest values of symmetry energy slope depends on the many-body forces parameter: 
the smaller the $\zeta$ parameter, the larger the symmetry  energy slope ~\cite{ROGomes2014}.
In particular, the smallest values of the symmetry energy slope for the parametrizations used in this work are those presented in Table~\ref{table:Table_coupling},
for each value of symmetry energy.

\begin{figure}
\begin{center}
 \begin{minipage}{.5\textwidth}
  \centering
  \includegraphics[width=1.1\linewidth]{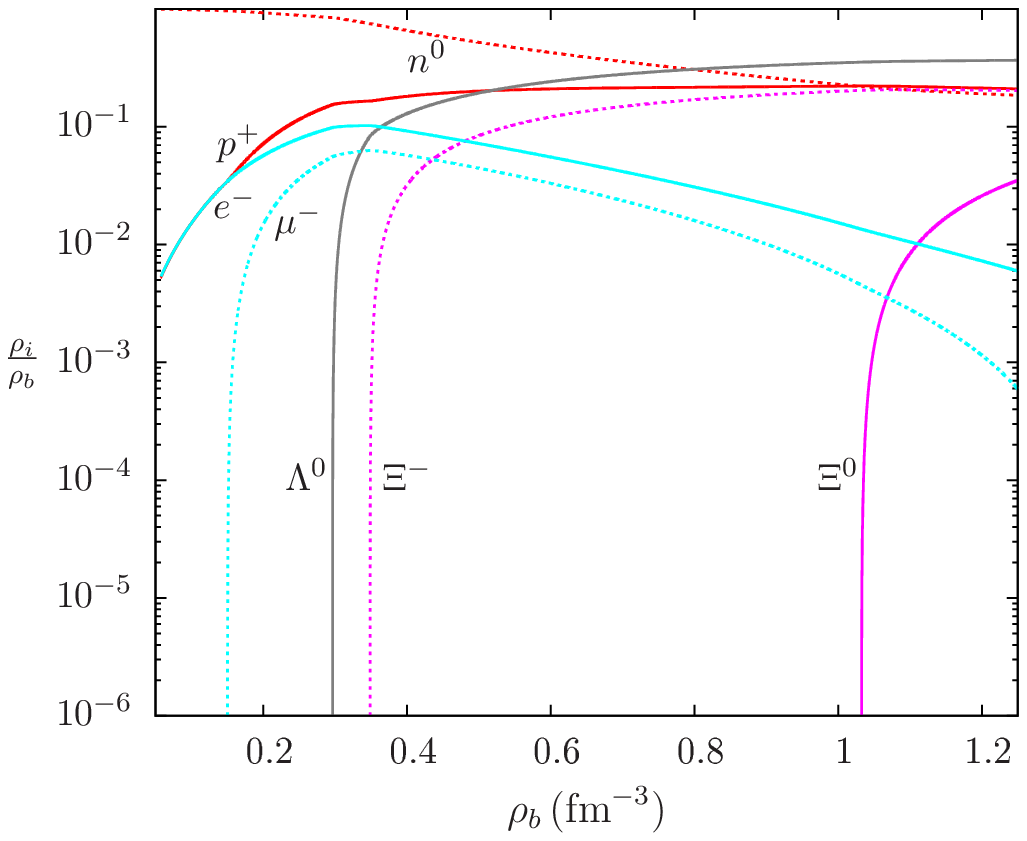}
  \end{minipage}%
\begin{minipage}{.5\textwidth}
  \centering
  \includegraphics[width=1.1\linewidth]{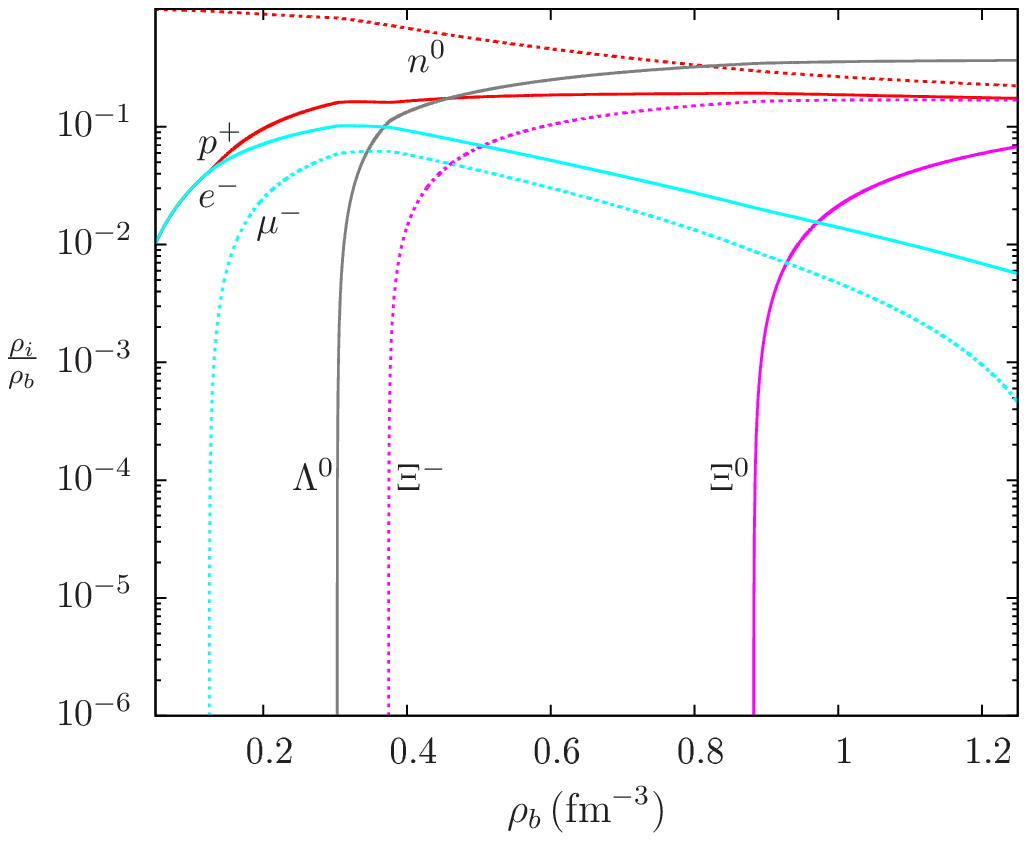}
  \end{minipage}
\caption{Particle population dependence on the symmetry energy $a_{sym}^0$, for the parametrization $\zeta=0.040$ and 
fixed symmetry energy slope $L_0=100$ MeV. 
The left panel shows the population for $a_{sym}^0=27$ MeV and the right for $a_{sym}^0=33$ MeV. The x-axis represents the 
baryon density and the y-axis the fraction of each particle species normalized by the total baryon density. }\label{pop_L100}
  \end{center}
\end{figure}

\begin{figure}
\begin{center}
\begin{minipage}{.5\textwidth}
  \centering
  \includegraphics[width=1.1\linewidth]{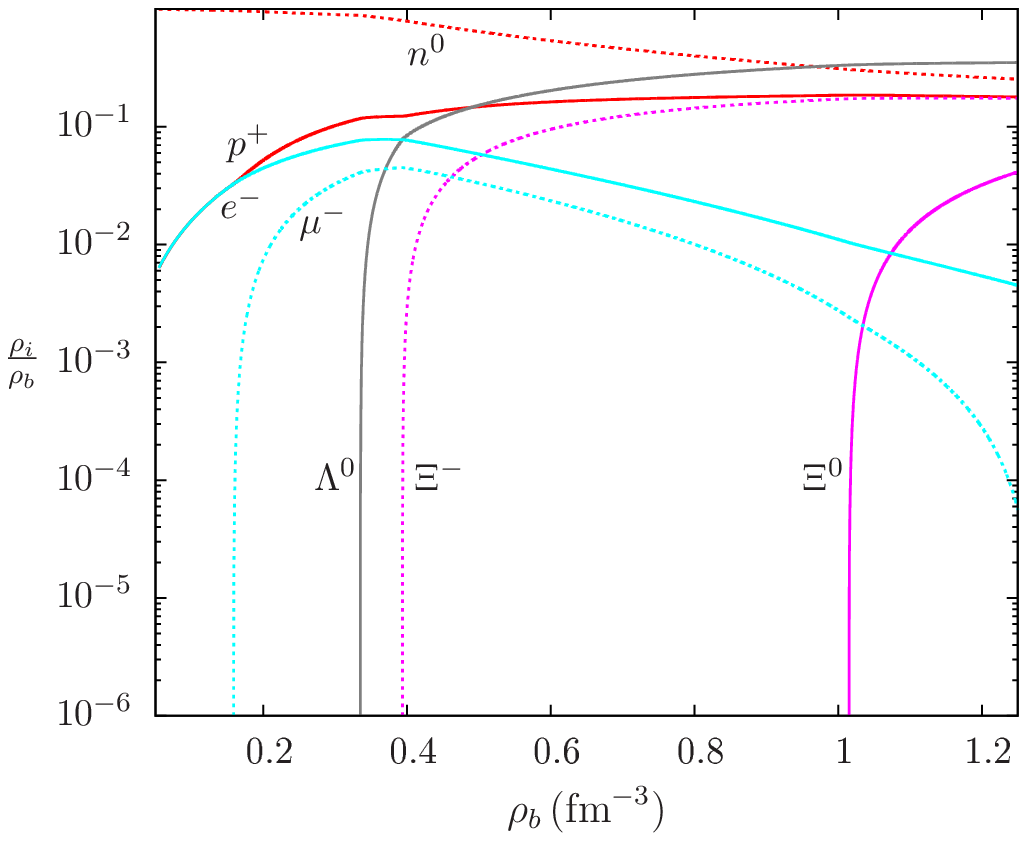}
  \end{minipage}%
\begin{minipage}{.5\textwidth}
  \centering
  \includegraphics[width=1.1\linewidth]{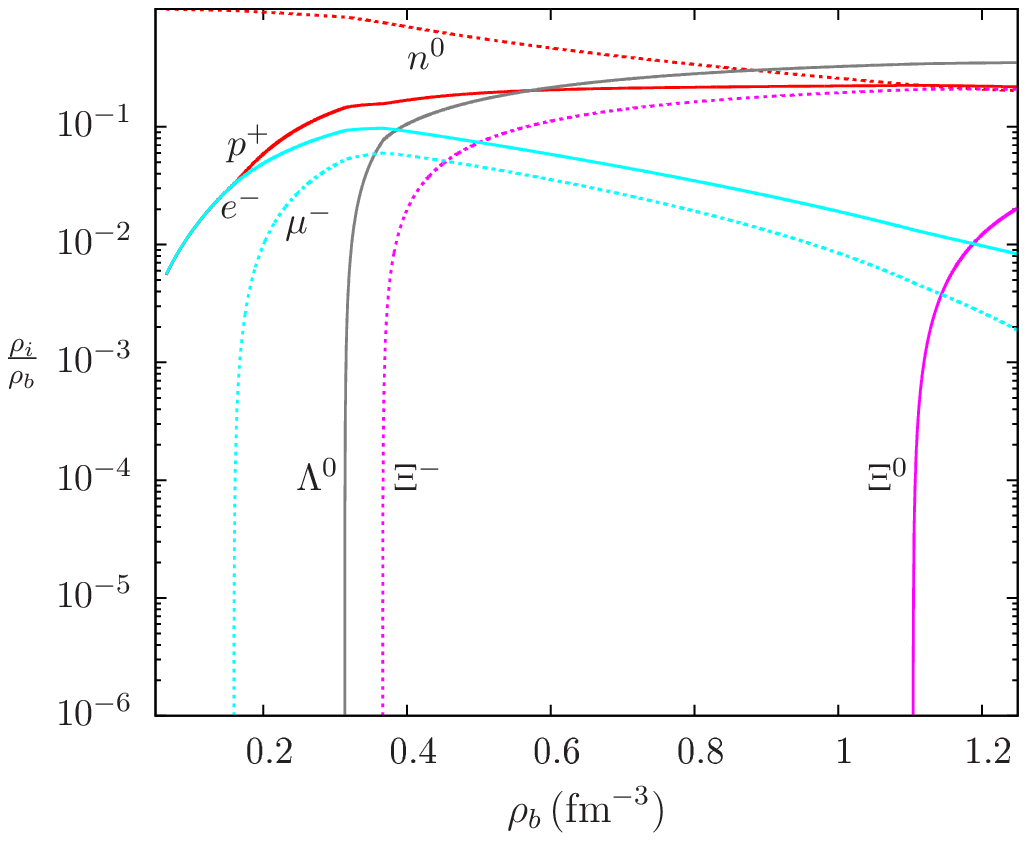}
  \end{minipage}
  \caption{Particle population dependence on the symmetry energy slope $L_{0}$, for the parametrization $\zeta=0.054$ and 
fixed symmetry energy $a_{sym}^0=25$ MeV. 
The left panel shows the population for $L_0=73$ MeV and the right for $L_0=90$ MeV. The axes are the same as Fig.~ \ref{pop_L100}.}\label{pop_a25}
\end{center}
\end{figure}

Fig.~\ref{pop_L100} and Fig.~\ref{pop_a25} show the particle population dependence on the symmetry energy and its slope, respectively.
In Fig.~\ref{pop_L100}, we show the results for the parametrizations $\zeta=0.040$ and $L_0= 100$ MeV, 
for  $a_{sym}^0=27$ MeV (left panel) and  $a_{sym}^0=33$ MeV (right panel). 
In Fig.~\ref{pop_a25}, we show the results for the parametrizations $\zeta=0.054$ and $a_{sym}^0=25$ MeV, 
for  $L_0=73$ MeV (left panel) and  $L_0=90$ MeV (right panel). 
The results show that lower values of the symmetry energy and higher values of the symmetry energy slope 
favor (disfavor) the appearence of particles of negative (positive) isospin.
Both quantities ($a_{sym}^0$ and $L_0$) have proportional effect on the coupling constants of the $\varrho$ and $\delta$ mesons,
i.e. lower values of $a_{sym}^0$ and $L_0$ provide weaker coupling of the baryon to these mesons.
These coupling constants, together with the isospin projection of the particles, 
play a role on the chemical potential and effective masses of baryons calculations and,
consequently, on the particle populations.

\begin{figure}
\begin{center}
\begin{minipage}{.5\textwidth}
  \centering
   \includegraphics[width=1.05\linewidth]{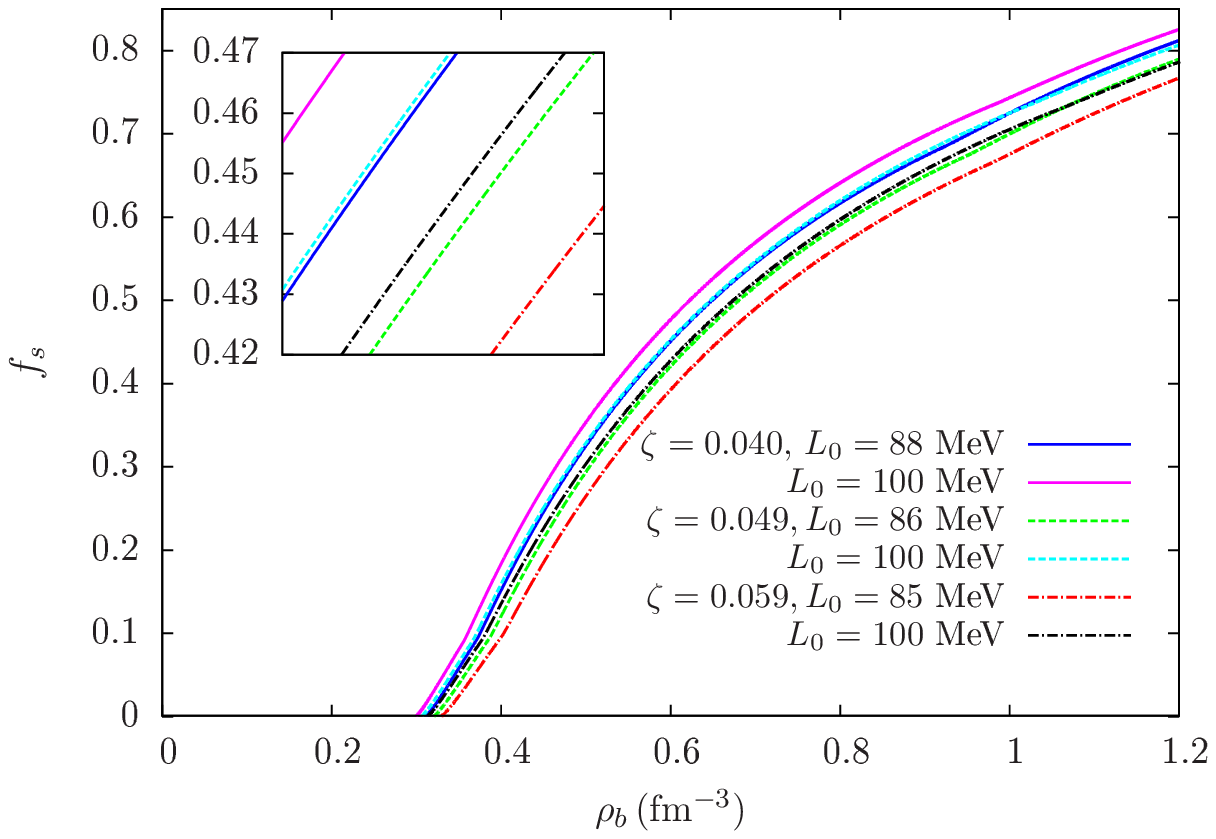}
  \end{minipage}%
\begin{minipage}{.5\textwidth}
  \centering
  \includegraphics[width=1.05\linewidth]{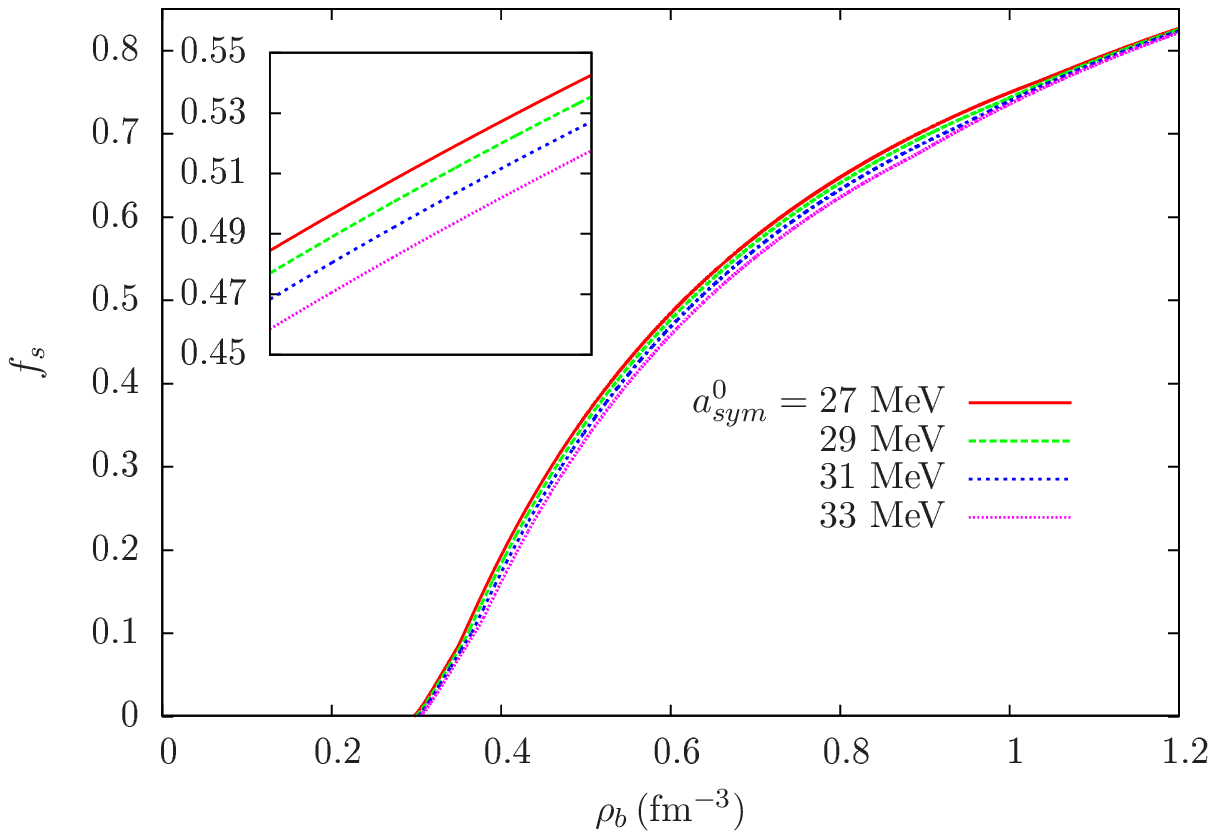}
  \end{minipage}
  \caption{\label{fs_Uc} Fraction of strangeness $f_s$ as a function of baryonic density. 
The left panel shows the results for three different parametrizations ($\zeta=0.040,\,0.049,\,0.059$)
and an interval of roughly $15$ MeV for the symmetry energy slope $L_0$, and fixed symmetry energy $a_{sym}^0=29$ MeV.
The right panel shows the strangeness results for the parametrization $\zeta=0.040$ and symmetry  energy slope $L_0=100$ MeV.}\label{fs}
  \end{center}
\end{figure}

Regarding the fraction of strangeness $fs$, the left panel on Fig.~\ref{fs_Uc} shows that both the many-body forces $\zeta$-parameter, as well as the 
slope of the symmetry energy have impact on hyperonic matter.
On the right panel of Fig.~\ref{fs_Uc}, one can also check that the effect of the symmetry energy on the fraction of strangeness is negligible, for 
the range of values studied in this work.

\begin{figure}
\begin{center}
\begin{minipage}{.5\textwidth}
  \centering
   \includegraphics[width=1.05\linewidth]{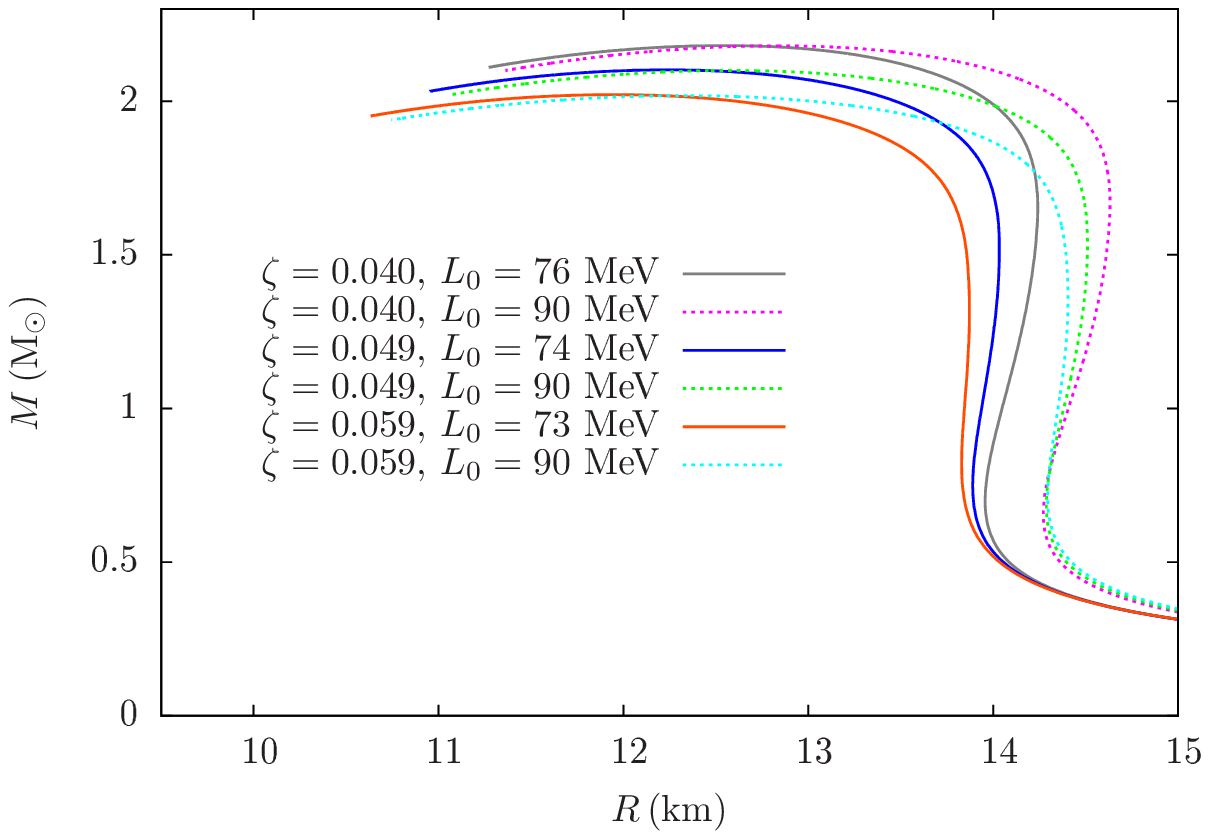}
  \end{minipage}%
\begin{minipage}{.5\textwidth}
  \centering
  \includegraphics[width=1.05\linewidth]{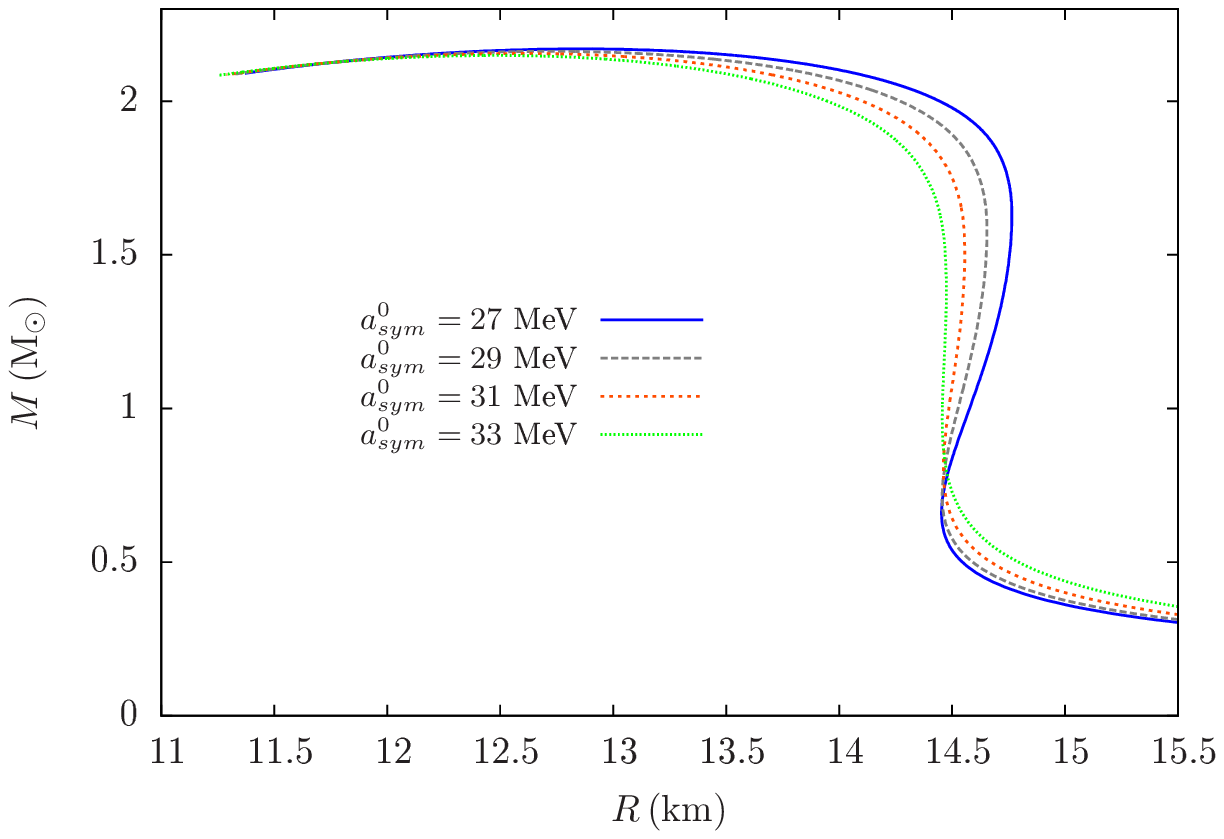}
  \end{minipage}
  \caption{\label{tov} Mass-radius relation for different values of symmetry energy and its slope. 
The left panel shows the results for three different parametrizations ($\zeta=0.040,\,0.049,\,0.059$)
and an interval of roughly $15$ MeV for the symmetry energy slope $L_0$, and fixed symmetry energy $a_{sym}^0=25$ MeV.
The right panel shows the strangeness results for the parametrization $\zeta=0.040$ and symmetry  energy slope $L_0=100$ MeV.}
  \end{center}
\end{figure}

Finaly, we apply our EoS to describe hyperon stars, solving the Tolman-Oppenheimer-Volkoff equations,
and perform the same analysis as above for the Mass-radius diagram in Fig. ~\ref{tov}.
On the left panel of Fig.~\ref{tov}, we show the results for three parametrizations ($\zeta=0.040,\,0.049,\,0.059$) and different values of $L_0$, 
for  $a_{sym}^0=29$ MeV. 
The results for the impact of the symmetry energy on the Mass-radius diagram are shown on the right panel of Fig.~\ref{tov}, 
for the parametrization $\zeta=0.040$ and $L_0=100$.
The results show that the symmetry energy and its slope have small effects on the maximum masses of the stars but,
on the other hand, have significant impact on the radii.
In particular, a shift of $~15$ MeV on the slope of the symmetry energy $L_0$ changes the radii of the canonical star ($1.4\,M_{\odot}$ star) in roughly $0.5$ km.
Similarly, a shift of $6$ MeV on the symmetry energy $a_{sym}^0$, changes the radii of the canonical star in $~0.4$ km.
Also, the many body forces parameter presents strong influence on both the maximum mass and radius of the stars,
with the highest maximum mass ($M_{max}=2.15M_{\odot}$) being provided by the parametrization $\zeta=0.040$~\cite{ROGomes2014}.

In this work we have analyzed the effects of many-body forces, symmetry energy and its slope on the properties of hyperonic matter and hyperon stars.
We have verified that the many-body forces contributions play an important role on the particle populations and on the macroscopic properties of hyperon stars.
Also, we showed that the asymmetric matter properties have non-negligigle influence on the particle populations and strangeness distribution of such systems, 
as well as have impact on the determination of the radii of hyperon stars, as already suggested by several studies ~\cite{Lopes:2014wda,ROGomes2014,Oertel2015}.  

\subsection*{Acknowledgement}

This work is partially supported by grant Nr. BEX 14116/13-8 of
the PDSE CAPES and Science without Borders programs which are an initiative of the Brazilian
Government.

\end{document}